\documentclass[a4paper,twocolumn,conference]{IEEEtran}
\usepackage[T1]{fontenc}
\usepackage{float}
\usepackage{amsthm}
\usepackage{graphicx}

\makeatletter

\pdfpageheight\paperheight
\pdfpagewidth\paperwidth

\floatstyle{ruled}
\newfloat{algorithm}{tbp}{loa}
\providecommand{\algorithmname}{Algorithm}
\floatname{algorithm}{\protect\algorithmname}

\theoremstyle{plain}

\theoremstyle{remark}
\newtheorem{rem}{\protect\remarkname}

\IEEEoverridecommandlockouts
\usepackage{ifpdf}
\usepackage{cite}
\ifCLASSOPTIONcompsoc
\usepackage[caption=false,font=normalsize,labelfont=sf,textfont=sf]{subfig}
\else
\usepackage[caption=false,font=footnotesize]{subfig}
\fi
\usepackage{amssymb}
\usepackage{amsmath}
\usepackage{float}
\usepackage{array}
\usepackage{makecell}

\makeatother

\providecommand{\remarkname}{Remark}
\providecommand{\theoremname}{Theorem}

\begin{document}

\title{Low-Complexity Transmission Mode Selection in MU-MIMO Systems}

\author{Haijing Liu,~Hui Gao,~Anzhong Hu, Tiejun~Lv\\Key Laboratory of
Trustworthy Distributed Computing and Service, Ministry of Education\\School
of Information and Communication Engineering\\Beijing University
of Posts and Telecommunications, Beijing, China 100876\\Email: \{Haijing\_LIU,
huigao, huanzhong, lvtiejun\}@bupt.edu.cn %
\thanks{This work is financially supported by the National Natural Science
Foundation of China (NSFC) under Grant No. 61271188.%
}}
\maketitle
\begin{abstract}
We propose a low-complexity transmission strategy in multi-user multiple-input
multiple-output downlink systems. The adaptive strategy adjusts the
precoding methods, denoted as the transmission mode, to improve the
system sum rates while maintaining the number of simultaneously served
users. Three linear precoding transmission modes are discussed, i.e.,
the block diagonalization zero-forcing, the cooperative zero-forcing
(CZF), and the cooperative matched-filter (CMF). Considering both
the number of data streams and the multiple-antenna configuration
of users, we modify the common CZF and CMF modes by allocating data
streams. Then, the transmission mode is selected between the modified
ones according to the asymptotic sum rate analyses. As instantaneous
channel state information is not needed for the mode selection, the
computational complexity is significantly reduced. Numerical simulations
confirm our analyses and demonstrate that the proposed scheme achieves
substantial performance gains with very low computational complexity.
\end{abstract}

\section{Introduction}

\setlength{\textfloatsep}{12pt plus 4pt minus 4pt}

Multi-user multiple-input multiple-output (MU-MIMO) systems have drawn
a lot of attention in the past decades. In order to handle the exponential
growth in mobile data traffic, many novel techniques such as mmWave,
full-duplex transmission and large-scale antenna system (LSAS) are
proposed to improve the performance of MU-MIMO systems. In particular,
LSAS, which achieves huge spectral-efficiency and energy-efficiency
gains \cite{su_investigation_2013,dai_spectrally_2013}, has recently
been considered as a technological breakthrough that holds great potential
for the future wireless communication.

In MU-MIMO downlink systems, the capacity region can be achieved by
employing Dirty Paper Coding (DPC) at the transmitter \cite{costa_writing_1983}.
But DPC is too complicated to be used in practice. Consequently, many
linear precoding schemes, which show impressive performance with much
lower complexity, have been proposed for practical multi-user downlink
systems, such as block diagonalization (BD) \cite{choi_transmit_2004},
zero-forcing (ZF), and matched-filter (MF). Recently, linear precoding
in LSAS has become a hot research topic. \cite{yang_performance_2013}
and \cite{lim_performance_2013} compared the performance of ZF and
MF precoders in multi-user multiple-input single-output (MU-MISO)
LSAS. \cite{wagner_large_2012} studied the regularized ZF (RZF) precoder
of MU-MISO LSAS in-depth, such as the optimal regularization parameter
for RZF. However, the existing studies on LSAS mainly focus on single-antenna
users rather than multi-antenna users, and have not been further developed
for transmission strategy design. Although various kinds of efficient
transmission strategies have been designed in the conventional MIMO
systems, they are not suitable for LSAS because most of them need
online computations relating to the instantaneous channel state information
(CSI). For example, \cite{love_multimode_2005} developed an efficient
transmission scheme by adapting linear precoding and signal modultaion,
relying upon the instantaneous channel capacity of probability of
error. With the increasing amount of system antennas, the computational
complexity of the above scheme becomes unbearable.

In this paper, we propose a low-complexity transmission strategy in
downlink MU-MIMO LSAS. The adaptive strategy adjusts the precoding
methods, denoted as the transmission mode, to enhance the system sum
rate performance. First, we get deterministic sum rate approximations
for the block diagonalization zero-forcing (BDZF), the cooperative
zero-forcing (CZF) and the cooperative matched-filter (CMF) modes.
In particular, we obtain a good upper bound of the sum rate of full-spatial-multiplexing
CZF, which has never been addressed in the aforementioned LSAS works.
These deterministic approximations enable us to propose an very low-complexity
transmission strategy without any instantaneous CSI. Therefore, in
order to achieve better sum rate performance and maintain the number
of simultaneously served users, we modify the common CZF and CMF precoding
schemes by scheduling the optimal amount of data streams as far as
possible. Furthermore, one of the modified CZF and CMF modes is selected
for higher sum rate. Such strategy cannot be developed in MU-MISO
works \cite{zhang_multi-mode_2011,yang_performance_2013,wagner_large_2012,lim_performance_2013}
as it takes advantage of the multi-antenna configuration of users.

\emph{Notations:} We use uppercase boldface letters for matrices
and lowercase boldface for vectors. $(\cdot)^H$, $(\cdot)^\dagger$,
$\textnormal{tr}(\cdot)$, $E[\cdot]$ and $\lfloor\cdot\rfloor$
denote the conjugate transpose, the pseudo-inverse, the trace, the
expectation, and the round down operation, respectively. $\mathcal{CN}(\mathbf{m},\mathbf{\Theta})$
denotes the circularly-symmetric complex Gaussian distribution with
mean vector $\mathbf{m}$ and covariance matrix $\mathbf{\Theta}$.
$\xrightarrow{a.s.}$ denotes the almost sure convergence, and $\xrightarrow{d}$
denotes convergence in distribution.

\section{System Model }

We consider a MU-MIMO downlink system composed of an $M$-antenna
base station and $K$ simultaneously served $N$-antenna users. We
assume $M\!\geq\!K$, so that user scheduling is not taken into account.
Perfect CSI is assumed available at the base station. The base station
sends $N_k$ data streams to the $k$-th user ($1\leq N_k\leq N$),
so that the total number of data streams of the system is $L=\sum^{K}_{k=1}N_k$.
The transmitted signal $\mathbf{x}\in\mathbb{C}^{M\!\times\!1}$ is
defined as \begin{equation}\label{eq11}\mathbf{x}=\mathbf{W}\mathbf{s}=\sum_{k=1}^{K}\mathbf{W}_k\mathbf{s}_k\end{equation}with
an average total power constraint $E[\textnormal{tr}(\mathbf{x}\mathbf{x}^H)]\leq P$,
where $P$ is the total available transmit power. $\mathbf{W}=[\mathbf{W}_1,\dots,\mathbf{W}_K]\in\mathbb{C}^{M\!\times\!L}$
is the total precoding matrix at the base station, and $\mathbf{s}=[\mathbf{s}_1^H,\dots,\mathbf{s}_K^H]^H\in\mathbb{C}^{L\!\times\!1}$
is the information-bearing vector from the base station to all the
$K$ users. $\mathbf{W}_k\in\mathbb{C}^{M\!\times\!N_k}$ and $\mathbf{s}_k\in\mathbb{C}^{N_k\!\times\!1}$
denotes the precoding matrix and the data vector for the $k$-th user,
respectively. $N_k$ antennas are pre-selected at the $k$-th user
to receive signals and the $N_k\!\times\!1$ received signal vector
is \begin{equation}\label{eq12}\mathbf{y}_k=\mathbf{H}_k\mathbf{x}+\mathbf{n}_k,\end{equation}
where $\mathbf{H}_k\in\mathbb{C}^{N_k\!\times\!M}$ with independent
$\mathcal{CN}(0,1)$ entries is the channel matrix from the base station
to the $k$-th user, $\mathbf{n}_k\in\mathbb{C}^{N_k\!\times\!1}$
with $\mathcal{CN}(0,\sigma^2_n)$ entries is the additive white Gaussian
noise at the $k$-th user. The $i$-th received data stream of the
$k$-th user (hereinafter referred to as data stream $(k,i),1\leq i\leq N_k,1\leq k\leq K$)
is given by \begin{equation}\label{eq13}\begin{split}y_{k,i}&=\mathbf{h}_{k,i}\mathbf{w}_{k,i}s_{k,i}\\&+\underbrace{\mathbf{h}_{k,i}\sum_{j=1,j\neq i}^{N_k}\mathbf{w}_{k,j}s_{k,j}}_{\textnormal{inter-stream interference}}+\underbrace{\mathbf{h}_{k,i}\sum_{l=1,l\neq k}^K\mathbf{W}_l\mathbf{s}_l}_{\textnormal{inter-user interference}}+n_{k,i},\end{split}\end{equation}where
$\mathbf{h}_{k,i}\in\mathbb{C}^{1\!\times\!M}$, $\mathbf{w}_{k,i}\in\mathbb{C}^{M\!\times\!1}$,
$s_{k,i}$ and $n_{k,i}$ is the $i$-th row of $\mathbf{H}_k$, the
$i$-th column of $\mathbf{W}_k$, the $i$-th element of $\mathbf{s}_{k}$
and the $i$-th entry of $\mathbf{n}_k$, respectively.

Denoting the signal-to-interference-and-noise ratio (SINR) of the
data stream $(k,i)$ by $\textnormal{SINR}_{k,i}$, the rate of data
stream $(k,i)$ is given by \begin{equation}\label{eq14}r_{k,i}=\log_2(1+\textnormal{SINR}_{k,i}).\end{equation}Then
the system sum rate can be calculated as \begin{equation}\label{eq15}\mathcal{R}=\sum_{k=1}^{K}\sum_{i=1}^{N_k}r_{k,i}.\end{equation}

\section{Transmission Modes Review}

We briefly go over the three transmission modes (i.e., BDZF, CZF and
CMF) of MU-MIMO in this section.

\subsection{BDZF}

In the BDZF mode, BD technique is utilized to precancel inter-user
interference followed by ZF precoders to remove the inter-stream interference
of each user. Hence, $\mathbf{W}_k$ is defined as a cascade of two
matrices, i.e., \begin{equation}\mathbf{W}_k=\alpha_k\mathbf{B}_k\mathbf{D}_k,\end{equation}
where $\alpha_k$ is the power control parameter. In this paper, uniform
power allocation among data streams is adopted, so we get $\alpha_k=\sqrt{PN_k/(L\text{tr}(\mathbf{B}_k\mathbf{D}_k\mathbf{D}_k^H\mathbf{B}_k^H))}$.
$\mathbf{B}_k\in\mathbb{C}^{M\!\times\!T_k}$ is designed with the
general method introduced in \cite{choi_transmit_2004} to remove
the inter-user interference in \eqref{eq13}. If we denote $\bar{\mathbf{H}}_k=\mathbf{H}_k\mathbf{B}_k$,
the ZF precoding matrix $\mathbf{D}_k\in\mathbb{C}^{T_k\!\times\!N_k}$
is $\mathbf{D}_k=\bar{\mathbf{H}}_k^\dagger$. To ensure the support
of $N_k$ data streams for the $k$-th user, $T_k$ should satisfy
the constraint $N_k\leq T_k\leq M+N_k-L$. The $\textnormal{SINR}$
of data stream $(k,i)$ is given by \begin{equation}\label{eq22}\textnormal{SINR}_{k,i}^{\text{BDZF}}=\frac{PN_k}{\sigma^2_{n}L\textnormal{tr}(\mathbf{B}_k\mathbf{D}_k\mathbf{D}_k^H\mathbf{B}_k^H)}=\frac{PN_k}{\sigma^2_{n}L\textnormal{tr}(\bar{\mathbf{H}}_k\bar{\mathbf{H}}_k^H)^{-1}}.\end{equation}

\subsection{CZF}

For CZF, the MU-MIMO system is treated as an equivalent single-user
MIMO (SU-MIMO) system. The equivalent channel $\mathbf{H}\in\mathbb{C}^{L\!\times\!M}$
from the base station to all the $K$ users is $\mathbf{H}=\left [\mathbf{H}_1^H,\mathbf{H}_2^H,\dots,\mathbf{H}_K^H\right ]^H$.
The CZF precoding matrix is \begin{equation}\mathbf{W}_{\textnormal{CZF}}=\beta(\mathbf{H})^\dagger=\beta\mathbf{H}^H(\mathbf{H}\mathbf{H}^H)^{-1},\end{equation}
where $\beta=\sqrt{P/\textnormal{tr}(\mathbf{H}\mathbf{H}^H)^{-1}}$
is utilized to normalize the transmit power. The SINR of data steam
$(k,i)$ is \begin{equation}\label{eq23}\textnormal{SINR}_{k,i}^{\text{CZF}}=\frac{P}{\sigma_{n}^2\textnormal{tr}(\mathbf{H}\mathbf{H}^H)^{-1}}.\end{equation}

\subsection{CMF}

For CMF, the MU-MIMO system is also treated as a SU-MIMO system. MF
instead of ZF precoding is utilized. The CMF precoding matrix is \begin{equation}\mathbf{W}_{\textnormal{CMF}}=\gamma\mathbf{H}^H,\end{equation}
where $\gamma=\sqrt{P/\text{tr}(\mathbf{H}\mathbf{H}^H)}$. For data
stream $(k,i)$, we get \begin{equation}y_{k,i}=\gamma\|\mathbf{h}_{k,i}\|^2s_{k,i}+\gamma\mathbf{h}_{k,i}\sum_{(l,m)\neq(k,i)}\mathbf{h}_{l,m}^H s_{l,m}+n_{k,i}.\end{equation}
The corresponding SINR is \begin{equation}\label{eq24}\textnormal{SINR}_{k,i}^{\text{CMF}}=\frac{\gamma^2\|\mathbf{h}_{k,i}\|^4}{\sigma^2_{n}+\gamma^2\mathbf{h}_{k,i}(\sum_{(l,m)\neq(k,i)}\mathbf{h}_{l,m}^H\mathbf{h}_{l,m})\mathbf{h}_{k,i}^H}.\end{equation}

\section{Proposed Transmission Mode Selection Scheme }

In this section, we propose a simple and effective transmission mode
selection scheme based on the analysis of the asymptotic sum rate
performance of the three aforementioned transmission modes.

\subsection{Sum Rate Analysis of Large-Scale MU-MIMO Systems}

In BDZF mode, in order to figure out the sum rate, we need to focus
on $\textnormal{tr}(\bar{\mathbf{H}}_k\bar{\mathbf{H}}_k^H)^{-1}$.
Since the elements of $\mathbf{H}_k$ are i.i.d. $\mathcal{CN}(0,1)$
random variables and $\mathbf{B}_k^H\mathbf{B}_k=\mathbf{I}_{N_k}$,
the elements of the equivalent channel $\bar{\mathbf{H}}_k$ are also
i.i.d. $\mathcal{CN}(0,1)$ random variables. Using the results of
\cite{tulino_random_2004}, we get $\textnormal{tr}(\bar{\mathbf{H}}_k\bar{\mathbf{H}}_k^H)^{-1}\xrightarrow{a.s.}N_k/(T_k-N_k)$
as $T_k$ and $N_k$ go to infinity while keeping a finite ratio $T_k/N_k$.
The sum rate is given by \begin{equation}\label{eq31}\mathcal{R}_{\textnormal{BDZF}}\xrightarrow{a.s.}\sum_{k=1}^{K}N_k\log_2\left(1+\frac{P(T_k-N_k)}{\sigma_n^2L}\right).\end{equation}

For CMF, similar to \cite{yang_performance_2013}, we have \begin{equation}\label{eq34}\mathcal{R}_{\textnormal{CMF}}\xrightarrow{d}L\log_2\left(1+\frac{P(M+1)}{\sigma^2_nL+P(L-1)}\right).\end{equation}

For CZF, in the case of $M>L$, as proposed in many previous works
\cite{tulino_random_2004,couillet_random_2011,wagner_large_2012,yang_performance_2013},
the sum rate is \begin{equation}\label{eq32}\mathcal{R}_{\textnormal{CZF}}\xrightarrow{a.s.}L\log_2\left(1+\frac{P(M-L)}{\sigma_n^2L}\right).\end{equation}In
the case of $M=L$, we propose a good upper bound for the sum rate
of CZF here. To the best of our knowledge, this has never been addressed
in the existing works. The sum rate can be rewritten as \begin{equation}\label{eq43}\begin{split}\mathcal{R}_{\textnormal{CZF}}&=L\log_2\left(1+\frac{P}{\sigma_n^2\text{tr}(\mathbf{H}\mathbf{H}^H)^{-1}}\right)\\&=L\log_2\left(1+\frac{P}{\sigma_n^2\sum_{l=1}^L\lambda_l^{-1}}\right),\end{split}\end{equation}where
$\lambda_l$ indicates the $l$-th eigenvalue of the matrix $\mathbf{HH}^H$.
At low signal-to-noise ratio (SNR), we have \begin{equation}\begin{split}\mathcal{R}_{\text{CZF}}&\approx\frac{PL}{\sigma_n^2\sum_{l=1}^L\lambda_l^{-1}}\\&\leq\sum_{l=1}^L\rho\lambda_l\\&\approx\sum_{l=1}^L\log_2(1+\rho\lambda_l),\end{split}\end{equation}where
$\rho\!=\!P/(L\sigma_n^2)$, ``$\approx$'' is derived from $\log_2(1+x)\approx x$
for sufficiently small $x$, and ``$\leq$'' is obtained by the
Arithmetic-Harmonic Mean inequality. At high SNR, we have \begin{equation}\begin{split}\mathcal{R}_{\text{CZF}}&\approx L\log_2\left(\frac{P}{\sigma_n^2\sum_{l=1}^L\lambda_l^{-1}}\right)\\&\leq\log_2(\prod_{l=1}^L\rho\lambda_l)\\&\approx\sum_{l=1}^L\log_2(1+\rho\lambda_l)\end{split}\end{equation}as
a result of $\log_2(1+x)\approx\log_2x$ for sufficiently large $x$,
and the less or equal relation is obtained by the Geometric-Harmonic
Mean inequality. Therefore, we obtain \begin{equation}\label{eq44}\mathcal{R}_{\text{CZF}}\leq\sum_{l=1}^L\log_2(1+\rho\lambda_l).\end{equation}
Similar to \cite{tulino_random_2004}, the upper bound of $\mathcal{R}_{\textnormal{CZF}}$
for $M=L$ is (we omit details here due to space limitations) \begin{equation}\label{eq39}\begin{split}&\mathcal{R}_{\textnormal{CZF}}^{\text{II}}\\&\xrightarrow{a.s.}2M\log_2\left(\frac{1\!+\!\sqrt{1\!+\!4\rho}}{2}\right)\!-\!\frac{M\log_2e}{4\rho}(\sqrt{1\!+\!4\rho}\!-\!1)^2.\end{split}\end{equation}

\subsection{Optimal Number of Data Streams in CZF}

When $M>L$, if $M$, $P$ and $\sigma_n^2$ are fixed, we can easily
get the explicit solution of the optimal number of data steams by
setting $d\mathcal{R}_{\textnormal{CZF}}/{dL}=0$, i.e., \begin{equation}\label{eq37}L^\text{I}_{\textnormal{CZF}}=\left\{\begin{aligned}&\frac{MP\omega}{(P-\sigma_n^2)(1+\omega)},&P\neq\sigma_n^2\\&\frac{M}{e},&P=\sigma_n^2\end{aligned},\right.\end{equation}where
$\omega$ is defined as $\omega=W\left((P-\sigma_n^2)/(\sigma_n^2e)\right)$
and $W(\cdot)$ is the Lambert W function.
\begin{rem}
Similar results about $L^1_{\text{CZF}}$ appear in \cite{couillet_random_2011,jung_optimal_2013}.
However, the case of $M=L$ should not be ignored. It is necessary
to discuss whether $M$ is the optimal number of data streams.
\end{rem}
With \eqref{eq37}, we can get the largest sum rate when $M>L$ as
\begin{equation}\label{eq38}\mathcal{R}^\text{I}_{\textnormal{CZF}}\xrightarrow{a.s.}\left\{\begin{aligned}&\frac{MP\omega}{(P-\sigma_n^2)\ln2},&P\neq\sigma_n^2\\&\frac{M}{e\ln2},&P=\sigma_n^2\end{aligned}\right..\end{equation}Hence,
by comparing the value of \eqref{eq39} and \eqref{eq38}, the optimal
number of data streams in CZF mode is given by \begin{equation}\label{eq40}L_{\textnormal{CZF}}^*=\left\{\begin{aligned}
&L^\text{I}_{\text{CZF}},&\mathcal{R}_{\textnormal{CZF}}^\text{I}\ge\mathcal{R}_{\textnormal{CZF}}^{\text{II}}\\
&M,&\mathcal{R}_{\textnormal{CZF}}^\text{I}<\mathcal{R}_{\textnormal{CZF}}^{\text{II}}
\end{aligned}\right..\end{equation}

Clearly, given $M,P$ and $\sigma_n^2$, we can easily get the optimal
number of data streams in terms of the sum rate.
\begin{rem}
$L_{\textnormal{CZF}}^*$ is usually smaller than $M$. If the transmit
power is large enough, we have $L_{\text{CZF}}^*=M$.
\end{rem}

\subsection{Modified CZF and CMF Schemes}

In the common CZF mode, we have $N_k=N$ (i.e., $L=NK$), which cannot
ensure the optimal sum rate performance as discussed above. Moreover,
$M/N$ limits the number of simultaneously served users in both CZF
and CMF modes.

We modify the common CZF mode through selecting the number of data
streams and allocating the data streams to each user. The configuration
$L=L^*_{\textnormal{CZF}}$ is ensured as far as possible to enhence
the system sum rate performance.

Algorithm 1 shows the details of the modified CZF scheme.

\begin{algorithm}[tbh]
\caption{Modified CZF Scheme}

\begin{enumerate}
\item {Calculate the optimal number of data streams $L^*_{\textnormal{CZF}}$ according to \eqref{eq40} with given $M,P \text{ and }  \sigma_n^2$.}
\item {Decide the number of data streams depending on $K$, $N$ and $L^*_{\textnormal{CZF}}$.}
\begin{description}
\item[A:]$L^*_{\text{CZF}}\geq NK$, each user utilizes all its $N$ antennas to receive data, i.e., $N_k=N$.
\item[B:]$K\leq L^*_{\text{CZF}}\leq NK$, set $N_k=\lfloor L^*_{\text{CZF}}/K\rfloor+\Delta_k$, where $\Delta_k=0\textnormal{ or }1$ is randomly chosen to satisfy $L=L^*_{\textnormal{CZF}}$.
\item[C:]$K>L^*_{\textnormal{CZF}}$, set $N_k=1$ for all $K$.
\end{description}
\item {The base station informs each user $N_k$ and transmits data streams to users with CZF precoding.}
\end{enumerate}
\end{algorithm}

In case B, taking advantage of the \textquotedblleft{}channel hardening
effect\textquotedblright{} of LSAS addressed in \cite{hochwald_multiple-antenna_2004},
we adopt the random selection of $\Delta_k$. In case C, one data
stream is delivered to each user to ensure the number of served users
while reducing the performance degradation as much as possible.

With the modified CZF scheme, the sum rate is \begin{equation}\label{eq41}\begin{split}
&\tilde{\mathcal{R}}_{\textnormal{CZF}}\xrightarrow{a.s.}\\
&\left\{\begin{aligned}
&NK\log_2\left(1+\frac{P(M-NK)}{\sigma_n^2NK}\right),1\leq K\leq\frac{L^*_{\textnormal{CZF}}}{N}\\
&\mathcal{R}_{\textnormal{CZF}}^*,\qquad\qquad\qquad\qquad\qquad\frac{L^*_{\textnormal{CZF}}}{N}<K\leq L^*_{\textnormal{CZF}}\\
&K\log_2\left(1+\frac{P(M-K)}{\sigma_n^2K}\right),\quad L^*_{\textnormal{CZF}}<K\leq M
\end{aligned}\right..
\end{split}\end{equation}

We also modify the common CMF scheme for its implementation in the
case of $M<KN$. In the case of $M\geq NK$, $N_k=N$ data streams
are sent to each user. When $M<KN$, we set $N_k=\lfloor M/K\rfloor+\Delta_k$,
where $\Delta_k$ is randomly set as 0 or 1 to ensure $L=M$. The
sum rate of the modified CMF mode is\begin{equation}\label{eq42}\begin{split}
&\tilde{\mathcal{R}}_{\textnormal{CMF}}\xrightarrow{a.s.}\\
&\left\{\begin{aligned}
&NK\log_2\left(1\!+\!\frac{P(M+1)}{\sigma^2_nNK+P(NK-1)}\right),&1\!\leq\!K\!\leq\frac{M}{N}&\\
&M\log_2\left(1\!+\!\frac{P(M+1)}{\sigma^2_nM+P(M-1)}\right),&\frac{M}{N}\!<\!K\!\leq\!M&
\end{aligned}\right..
\end{split}\end{equation}

\subsection{Transmission Mode Selection According to $K$\label{sub:Mode-Selection-According}}

In BDZF mode, apparently, the sum rate is maximized in the case of
$T_k=M+N_k-L$. So the largest sum rate is \begin{equation}\label{eq35}\mathcal{R}_\textnormal{BDZF}^{\textnormal{MAX}}\xrightarrow{a.s.}L\log_2\left(1+\frac{P(M-L)}{\sigma_n^2L}\right),\end{equation}which
is same as \eqref{eq32}. Moreover, $K$ SVD and $K$ pseudo-inverse
operations are required to find a precoding matrix $\mathbf{W}$ in
BDZF mode, while only one pseudo-inverse is needed in CZF mode. Namely,
CZF mode can achieve the same or better sum rate performance than
BDZF with a significantly lower computational complexity. Consequently,
we only consider the modified CZF and CMF modes in the proposed mode
selection scheme.

Given the system parameters $M,N,K,P$ and $\sigma_n^2$, we select
the transmission mode which provides higher system sum rate. The modified
CZF mode is selected for data transmission in the case of $\tilde{\mathcal{R}}_{\textnormal{CZF}}\geq\tilde{\mathcal{R}}_{\textnormal{CMF}}$.
In another case (i.e., $\tilde{\mathcal{R}}_{\textnormal{CZF}}<\tilde{\mathcal{R}}_{\textnormal{CMF}}$)
the modified CMF mode is selected.

Furthermore, as the antenna configuration of the base station and
users, the total available transmit power and the thermal noise are
almost unchanged in practical cellular systems, i.e., $M,N,P$ and
$\sigma_n^2$ are fixed, we can pre-calculate sum rates with \eqref{eq41}
and \eqref{eq42} for various $K$, and then find the intervals of
$K$ for CZF and CMF modes in advance. Hence, the proposed transmission
mode selection only depends on $K$. The details are presented in
Algorithm 2.

\begin{algorithm}[tbh]
\caption{Proposed Transmission Mode Selection Scheme}

\begin{enumerate}
\item {Find $L^*_{\textnormal{CZF}}$ according to \eqref{eq37}, \eqref{eq38}, \eqref{eq39} and \eqref{eq40} with given $M,N,P$ and $\sigma_n^2$.}
\item {Find the intervals of $K$ for different transmission modes as follows:\begin{equation}\begin{split}\pi_{\textnormal{CZF}}&=\{K|K\in\mathbb{N}^+,\tilde{\mathcal{R}}_{\textnormal{CZF}}\geq\tilde{\mathcal{R}}_{\textnormal{CMF}}\},\\\pi_{\textnormal{CMF}}&=\{K|K\in\mathbb{N}^+,\tilde{\mathcal{R}}_{\textnormal{CZF}}<\tilde{\mathcal{R}}_{\textnormal{CMF}}\},\end{split}\end{equation}where $\tilde{\mathcal{R}}_{\textnormal{CZF}}$ and $\tilde{\mathcal{R}}_{\textnormal{CMF}}$ are calculated as \eqref{eq41} and \eqref{eq42}, respectively. }
\item {Select the modified CZF or CMF mode for data transmission according to $K$.}
\end{enumerate}
\end{algorithm}

\begin{rem}
With the modified CZF and CMF schemes, besides better sum rate performance
(especially for a large number $K$ of users), the proposed scheme
supports up to $K\!=\!M$ users simultaneously compared with $K\!=\!M/N$
in the common schemes.
\end{rem}

\subsection{Complexity Analysis}

For the system model that we discuss in this paper, if we select the
transmission by the brute force searching, about $\left(\binom{1}{N}+\cdots+\binom{N}{N}\right)^K=(2^N-1)^K$
data stram allocations need to be checked for finding the best mode,
which is unacceptable in practical implementations. In contrast, in
the proposed scheme, we only need to determine the interval ($\pi_{\text{CZF}}$
or $\pi_{\text{CMF}}$) that $K$ belongs to during the selection.
The intervals can be solved by any standard numerical methods, and
need to be updated only when $M,N,P \text{ or }\sigma_n^2$ changes.
If the system configuration is unchanged, only Step 3) in Algorithm
2 is required for the selection.

\section{Numerical Simulations}

This section illustrates the performance of the proposed transmission
mode selection scheme by Monte Carlo simulations. The transmit SNR
is defined as $\text{SNR}=P/\sigma_n^2$.

\begin{figure}[t]
\centering

\includegraphics[width=8cm]{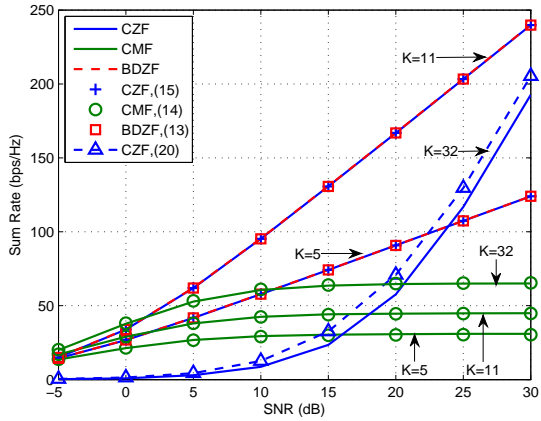}\caption{Sum rates for $K=5,11,32$ in various transmission modes with $M=64,N=2$. }

\centering{}\label{FIG 2}
\end{figure}

In Fig. \ref{FIG 2}, we discuss an $M\!=\!64,N\!=\!2$ MU-MIMO system
and set $T_k\!=\!M\!-\!N_k\!+\!L$ in BDZF mode. The sum rate approximations
of the BDZF, CZF and CMF modes are compared with the ergodic sum rates
averaged over 10000 independent channel realizations. It can be observed
that the approximations in \eqref{eq31}, \eqref{eq32} and \eqref{eq34}
are accurate for $M/L>1$. \eqref{eq39} is also an effective upper
bound of $\mathcal{R}_{\text{CZF}}$ in the case of $M/L=1$. As mentioned
in Section \ref{sub:Mode-Selection-According}, the BDZF mode has
almost the same sum rate performance as the CZF mode. Furthermore,
it is clear the CZF and BDZF curves for $K=11$ are above those for
$K=5$ and $K=32$, which confirms that the sum rates of CZF and BDZF
are not monotonically increasing with the increasing $K$ under certain
SNR conditions. However, the sum rate of CMF increases with the growth
of $K$. In addition, the CMF mode achieves better sum rate performance
than the CZF and the BDZF modes at low SNR while the opposite is true
at high SNR.

\begin{figure}[t]
\centering

\includegraphics[width=8cm]{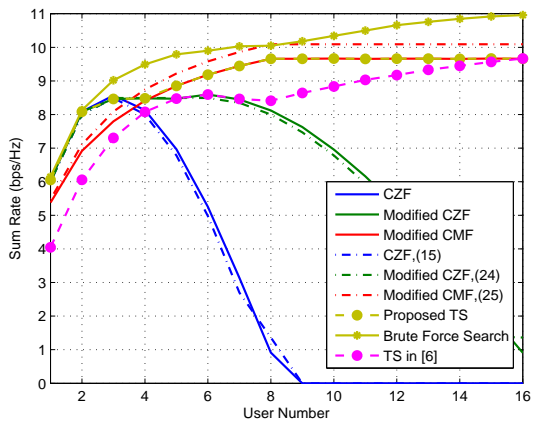}\caption{Sum rates in various transmission modes with $\text{SNR}=0\text{ dB},M=16$.}

\label{FIG 4}
\end{figure}

The sum rates of various transmission modes with $M=16,N=2$ are presented
in Fig. \ref{FIG 4}. The transmit SNR is 0 dB and the power is uniformly
allocated on each data stream. The legend ``Proposed TS'' indicates
the proposed transmission mode selection scheme and the blue curves
show the performance of the common CZF scheme. As expected, the proposed
scheme combines the advantages of CZF and CMF, and supports up to
16 users at the same time, while the common modes only support 8 users.
The brute force search is utilized as a benchmark to evaluate the
sum rate performance. It can be seen that about 90\% sum rates of
the brute force search scheme are achieved by the proposed scheme.
We also show the performance of the transmission selection scheme
proposed in \cite{lim_performance_2013} for $M=16,N=1$. It is obviously
that for the same number $K$ ($K<M$) of served users, our proposed
scheme with multi-antenna users obtains better sum rate performance.

\section{Conclusions}

We have proposed a low-complexity transmission scheme in MU-MIMO systems
in the paper. Based on the comprehensive discussions on the sum rate
performance of the BDZF, CZF and CMF precoding schemes, we have modified
the common CZF and CMF schemes and developed a novel transmission
mode selection approach to enhance the system sum rate performance
while maintaining the number of simultaneously served users. The simulations
show that our proposed transmission mode selection scheme achieves
near optimal sum rate performance with extremely low computational
complexity. In the future, we will focus on the low-complexity transmission
mode selection scheme in multi-cell large-scale downlink systems.

\bibliographystyle{IEEEtran}
\bibliography{TS}

\begin{thebibliography}{10}
\providecommand{\url}[1]{#1}
\csname url@samestyle\endcsname
\providecommand{\newblock}{\relax}
\providecommand{\bibinfo}[2]{#2}
\providecommand{\BIBentrySTDinterwordspacing}{\spaceskip=0pt\relax}
\providecommand{\BIBentryALTinterwordstretchfactor}{4}
\providecommand{\BIBentryALTinterwordspacing}{\spaceskip=\fontdimen2\font plus
\BIBentryALTinterwordstretchfactor\fontdimen3\font minus
  \fontdimen4\font\relax}
\providecommand{\BIBforeignlanguage}[2]{{%
\expandafter\ifx\csname l@#1\endcsname\relax
\typeout{** WARNING: IEEEtran.bst: No hyphenation pattern has been}%
\typeout{** loaded for the language `#1'. Using the pattern for}%
\typeout{** the default language instead.}%
\else
\language=\csname l@#1\endcsname
\fi
#2}}
\providecommand{\BIBdecl}{\relax}
\BIBdecl

\bibitem{su_investigation_2013}
X.~Su, J.~Zeng, L.-P. Rong, and Y.-J. Kuang, ``Investigation on key
  technologies in large-scale {MIMO},'' \emph{Journal of Computer Science and
  Technology}, vol.~28, no.~3, pp. 412--419, May 2013.

\bibitem{dai_spectrally_2013}
L.~Dai, Z.~Wang, and Z.~Yang, ``Spectrally efficient time-frequency training
  {OFDM} for mobile large-scale {MIMO} systems,'' \emph{{IEEE} J. Sel. Areas
  Commun.}, vol.~31, no.~2, pp. 251--263, Feb. 2013.

\bibitem{costa_writing_1983}
M.~H.~M. Costa, ``Writing on dirty paper,'' \emph{{IEEE} Trans. Inf. Theory},
  vol.~29, pp. 439--441, May 1983.

\bibitem{choi_transmit_2004}
L.-U. Choi and R.~Murch, ``A transmit preprocessing technique for multiuser
  {MIMO} systems using a decomposition approach,'' \emph{{IEEE} Trans. Wireless
  Commun.}, vol.~3, pp. 20--24, Jan. 2004.

\bibitem{yang_performance_2013}
H.~Yang and T.~Marzetta, ``Performance of conjugate and zero-forcing
  beamforming in large-scale antenna systems,'' \emph{{IEEE} J. Sel. Areas
  Commun.}, vol.~31, pp. 172--179, Feb. 2013.

\bibitem{lim_performance_2013}
\BIBentryALTinterwordspacing
Y.-G. Lim, C.-B. Chae, and G.~Caire, ``Performance analysis of massive {MIMO}
  for cell-boundary users,'' {arXiv} e-print 1309.7817, Sep. 2013. [Online].
  Available: \url{http://arxiv.org/abs/1309.7817}
\BIBentrySTDinterwordspacing

\bibitem{wagner_large_2012}
S.~Wagner, R.~Couillet, M.~Debbah, and D.~T.~M. Slock, ``Large system analysis
  of linear precoding in correlated {MISO} broadcast channels under limited
  feedback,'' \emph{{IEEE} Trans. Inf. Theory}, vol.~58, pp. 4509--4537, Jul.
  2012.

\bibitem{love_multimode_2005}
D.~Love and R.~Heath, ``Multimode precoding for {MIMO} wireless systems,''
  \emph{{IEEE} Trans. Signal Process.}, vol.~53, pp. 3674--3687, Oct. 2005.

\bibitem{zhang_multi-mode_2011}
J.~Zhang, M.~Kountouris, J.~Andrews, and R.~Heath, ``Multi-mode transmission
  for the {MIMO} broadcast channel with imperfect channel state information,''
  \emph{{IEEE} Trans. Commun.}, vol.~59, pp. 803--814, Mar. 2011.

\bibitem{tulino_random_2004}
A.~M. Tulino and S.~Verd$\acute{\text{u}}$,
  \emph{\BIBforeignlanguage{en}{Random Matrix Theory and Wireless
  Communications}}, 1st~ed.\hskip 1em plus 0.5em minus 0.4em\relax Hanover: Now
  Publishers Inc, 2004, pp. 14--15.

\bibitem{couillet_random_2011}
R.~Couillet and M.~Debbah, \emph{\BIBforeignlanguage{en}{Random Matrix Methods
  for Wireless Communications}}, 1st~ed.\hskip 1em plus 0.5em minus 0.4em\relax
  Cambridge, United Kingdom: CUP, 2011.

\bibitem{jung_optimal_2013}
M.~Jung, Y.~Kim, J.~Lee, and S.~Choi, ``Optimal number of users in zero-forcing
  based multiuser {MIMO} systems with large number of antennas,'' \emph{J.
  Commun. and Networks}, vol.~15, no.~4, pp. 362--369, Apr. 2013.

\bibitem{hochwald_multiple-antenna_2004}
B.~Hochwald, T.~Marzetta, and V.~Tarokh, ``Multiple-antenna channel hardening
  and its implications for rate feedback and scheduling,'' \emph{{IEEE} Trans.
  Inf. Theory}, vol.~50, pp. 1893--1909, Sep. 2004.

\end{thebibliography}

\end{document}